%% file: main.tex
\documentclass[conference]{IEEEtran}
\IEEEoverridecommandlockouts
\usepackage{cite}
\usepackage{amsmath,amssymb,amsfonts}
\usepackage{algorithmic}
\usepackage{graphicx}
\usepackage{textcomp}
\usepackage{xcolor}
\usepackage{url}
\usepackage{subcaption}
\def\BibTeX{{\rm B\kern-.05em{\sc i\kern-.025em b}\kern-.08em
    T\kern-.1667em\lower.7ex\hbox{E}\kern-.125emX}}
\begin{document}

\title{ Deep Frequency Attention Networks for Single Snapshot Sparse Array Interpolation\\
\thanks{This work was supported in part by U.S. National Science
Foundation (NSF) under Grants CCF-2153386 and ECCS-2340029.}
}
\author{
	\IEEEauthorblockN{Ruxin~Zheng, Shunqiao Sun and Hongshan~Liu}\\
\IEEEauthorblockA{Department of Electrical and Computer Engineering, The University of Alabama, Tuscaloosa, AL, USA\\	
E-mails: \texttt{\{rzheng9,hliu75\}@crimson.ua.edu, shunqiao.sun@ua.edu}}
}

\maketitle

\input{section/0_abstract}
\input{section/1_intro}
\input{section/2_system_model}

\input{section/3_proposed}

\input{section/4_performance}
\input{section/5_conclusion}
\bibliographystyle{IEEEtran}
\bibliography{refs}
\end{document}

%% file: section/0_abstract.tex
\begin{abstract}
Sparse arrays have been widely exploited in radar systems because of their advantages in achieving large array aperture at low hardware cost, while significantly reducing mutual coupling. However, sparse arrays suffer from high sidelobes which may lead to false detections. 
Missing elements in sparse arrays can be interpolated using the sparse array measurements. In snapshot-limited scenarios, such as automotive radar, it is challenging to utilize difference coarrays which require a large number of snapshots to construct a covariance matrix for interpolation. For single snapshot sparse array interpolation, traditional model-based methods, while effective, require expert knowledge for hyperparameter tuning, lack task-specific adaptability, and incur high computational costs. In this paper, we propose a novel deep learning-based single snapshot sparse array interpolation network that addresses these challenges by leveraging a frequency-domain attention mechanism. The proposed approach transforms the sparse signal into the frequency domain, where the attention mechanism focuses on key spectral regions, enabling improved interpolation of missing elements even in low signal-to-noise ratio (SNR) conditions. By minimizing computational costs and enhancing interpolation accuracy, the proposed method demonstrates superior performance compared to traditional approaches, making it well-suited for automotive radar applications.

\end{abstract}


%% file: section/1_intro.tex
\section{Introduction}

Radar technology is integral to autonomous driving, offering robust performance in adverse weather conditions \cite{SUN_SPM_Feature_Article_2020, Ruxin_TAES_2023,xu2023automotive}. Automotive radar must provide high-resolution four-dimensional data, including range, Doppler shifts, azimuth, and elevation angles, while remaining cost-effective for mass production \cite{Sun_JSTSP_2021}. While range and Doppler resolution are determined by the waveform bandwidth and the coherent processing interval, improving angular resolution is essential for precise localization and tracking. MIMO radar, which has become the automotive industry standard, enhances angular resolution by expanding the virtual array aperture beyond the physical antenna dimensions \cite{SUN_SPM_Feature_Article_2020}. This capability can be further improved using super-resolution direction of arrival (DOA) estimation methods.

Achieving large antenna apertures for higher angular resolution is challenging, especially for filled arrays that require a significant number of elements. Sparse arrays offer an efficient and cost-effective alternative by enabling larger apertures with fewer elements and reducing mutual coupling \cite{SUN_ICASSP_2020, Sun_JSTSP_2021,Wei_Liu_sparse_array_automotive_TSP_2024,Sparse_array_EuRAD_2019,Lifan_sparse_array_Asilomar_2023}. However, the design of optimal sparse arrays remains complex, since the ideal configuration depends on various requirements, making a universal solution impractical \cite{Zheng2023asilomarSparse, lin2022design}. 

Missing elements in sparse arrays can elevate sidelobe levels, making it challenging for DOA estimation. In snapshot-limited scenarios, such as automotive radar, the use of difference coarrays \cite{Vaidyanathan_Coprime_2011,qin2015generalized} becomes challenging due to the need for a large number of snapshots to construct a covariance matrix for interpolation. Hankel matrix completion methods \cite{Shuimei_TSP_2021,zhang2020doa,SUN_Multi_Freq_Asilomar_2020,cai2019fast} work for single snapshot sparse array interpolation by exploiting low-rank properties. However, model-based matrix completion techniques impose high computational costs due to singular value decomposition (SVD) operations \cite{cai2010singular}.

Deep neural networks have demonstrated superior performance in low-rank matrix completion \cite{fan2017deep,hu2024iht}, yet existing models suffer from excessive depth, large parameter counts, and a fundamental limitation: they are designed for fixed sparse array geometries. Adapting to varying array structures without retraining remains a key challenge\cite{zheng2024antenna}. Additionally, ensuring resilience against random sensor failures \cite{4286013} is crucial for maintaining radar system reliability.

In this paper, we propose a novel deep learning framework for single snapshot sparse array interpolation that incorporates a sparse and noise signal augmentation module. This module randomly masks input signals to simulate various sparse array structures while introducing noise, enhancing the model's adaptability. Furthermore, our approach tokenizes the signal in the frequency domain and leverages an attention mechanism \cite{vaswani2017attention} to refine interpolation performance. Comprehensive experiments demonstrate the framework's robustness across different sparse array configurations, providing a reliable solution for automotive radar systems. By improving generalizability and resilience, this work advances the state of sparse array interpolation and contributes a novel perspective to the field.

%% file: section/2_system_model.tex
\section{Signal Model}

\begin{figure*}[ht]
\centering
\includegraphics[width=0.75 \linewidth]{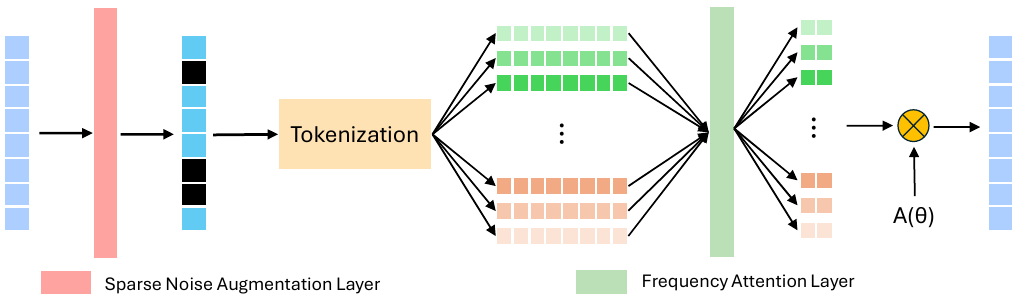}

\caption{Network architecture incorporating a sparse noise augmentation layer and a frequency attention mechanism within a signal reconstruction framework.
}
\label{arch}
\vspace{-5mm}
\end{figure*}

Consider a general uniform linear antenna array (ULA) with \( N \) elements observing \( K \) far-field point targets located at angles \( \theta_k \) for \( k = 1, \dots, K \). The received single snapshot array response is modeled as  
\begin{align}\label{sig}  
\mathbf{y} = \mathbf{A}(\boldsymbol{\theta}) \mathbf{s} + \mathbf{n},  
\end{align}  
where \( \mathbf{n} \) represents spatially and temporally white noise, and $\mathbf{A}(\boldsymbol{\theta}) $ is the \( N \times K \) array manifold matrix given by  
\begin{align}  
\mathbf{A}(\boldsymbol{\theta}) = \left[ \mathbf{a}(\theta_1), \mathbf{a}(\theta_2), \dots, \mathbf{a}(\theta_K) \right],  
\end{align}  
where $\mathbf{a}(\theta) = \left[ 1, e^{\frac{2\pi d_2}{\lambda} \sin{\theta}}, \dots, e^{\frac{2\pi d_N}{\lambda} \sin{\theta}} \right]^T $ is the array steering vector,
and \( d_n \) denotes the spacing between the \( n \)-th element and the reference element, and \( \mathbf{s} = [s_1, s_2, \dots, s_K]^T \) represents the transmitted source waveform vector.  

Sparse arrays offer a practical approach to direction finding by strategically reducing the number of active antenna elements while preserving the overall aperture. 
A sparse array consists of a subset of the full array elements, which means that only measurements from selected sensor positions are available. Let \( \mathbf{M} \) be an \( M \times N \) binary selection matrix (\( M < N \)) that maps the full array response to the observed sparse measurements. The single snapshot sparse array received signal is
\begin{align}  
\mathbf{y}_s = \mathbf{M} \mathbf{y},  
\end{align}  
where \( \mathbf{y}_s \) represents the observed sparse measurements, obtained by selecting a subset of the elements from \( \mathbf{y} \), and the noise term in the sparse array is inherently included in the selection process. The sparsity level of an sparse array is 
\begin{equation}
\text{Sparsity} = 1 - \frac{M}{N},
\end{equation}
which quantifies the proportion of removed elements relative to the full array.

In this paper, our goal is to reconstruct \( \mathbf{A}(\theta) \mathbf{s} \) from \( \mathbf{y}_s \), recovering the full signal while suppressing noise.

%% file: section/3_proposed.tex
\section{Frequency Attention Network for Signal Reconstruction}
In this section, we present the Frequency Attention Network (FA-Net), designed to reconstruct the original signal from a sparse array while effectively denoising it. Additionally, we detail the process of training data generation and outline the training methodology employed for the network.

\subsection{The FA-Net Architecture}
The FA-Net is designed with three key components: a sparse and noise augmentation layer, frequency domain tokenization, and a frequency domain attention mechanism, as illustrated in Fig. \ref{arch}. The process begins by introducing noise and randomly masking elements in the original data to simulate sparsity. The data is then transformed into the frequency domain, where it undergoes tokenization to facilitate structured processing. Finally, a frequency domain attention mechanism is applied to assign importance scores to different frequency components, enabling the network to distinguish true signal information from noise and improve reconstruction accuracy.

\subsubsection{Sparse and Noise Augmentation Layer}

Data augmentation enhances model generalization and mitigates overfitting \cite{shorten2019survey}. In computer vision, transformations like flipping and rotation improve robustness, while in signal processing, the sparse noise augmentation layer introduces structured sparsity and controlled noise for better signal reconstruction. This layer applies a random binary mask to the input signal, with sparsity controlled by a configurable maximum sparsity parameter. Gaussian noise is added, with its magnitude regulated by a specified SNR range. During training, the number of masked elements is randomly selected per batch, and noise is applied with an SNR sampled from the defined range.

During training, the number of active antennas is dynamically determined by this augmentation layer, ensuring diverse training scenarios. However, during evaluation, the sparsity pattern is fixed using predefined thresholding algorithms to maintain consistency. This sparse noise augmentation layer is only applied during training to enhance the network’s robustness in handling incomplete and noisy signals.
\begin{figure}[ht]
\centering
\includegraphics[width=0.9 \linewidth]{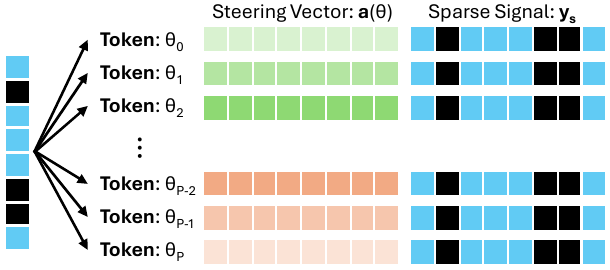}
\vspace{-1mm}
\caption{Frequency domain tokenization.
}
\label{token}
\end{figure}

\subsubsection{Frequency Domain Tokenization}
Random sparsity results in missing signal elements, posing challenges for implementing a deep learning network capable of robust signal reconstruction. To address this, we transform the signal into the frequency domain through a tokenization process. Specifically, we define a frequency grid of size P, where each frequency bin corresponds to a unique token, as illustrated in Fig. \ref{token}. For each token associated with $\theta_p$, the steering vector $\mathbf{a}(\theta_p)$ is concatenated with the sparse signal, allowing the network to effectively capture both spatial and spectral information.

\subsubsection{Attention Mechanism}
In signal reconstruction from sparse arrays, different frequency components contribute unequally to the signal structure. Some frequencies carry essential target information, while others are more affected by noise, distortions, or artifacts from sparse array geometry. The irregular element spacing in sparse arrays leads to non-uniform beam patterns, causing certain frequencies to be more impacted by sidelobes or missing data. To address this, an attention mechanism is introduced to selectively emphasize important frequency components while reducing the influence of less reliable ones.

\begin{figure}[ht]
\centering
\includegraphics[width= 0.95\linewidth]{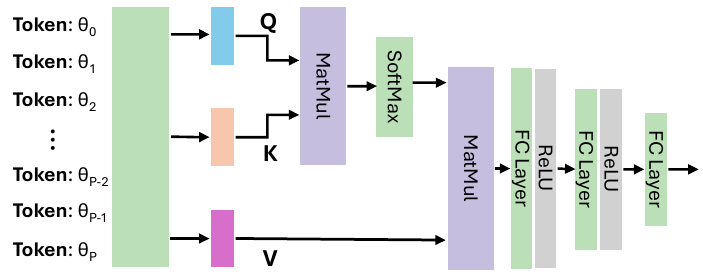}
\vspace{-1mm}
\caption{Attention layer.
}
\label{attn}
\end{figure}

As shown in Fig.~\ref{attn}, the attention mechanism learns frequency-domain dependencies by dynamically adjusting the contribution of each frequency token. Each frequency bin \( \theta_p \) is treated as a token that combines spatial and spectral information. These tokens are mapped to query (\( \mathbf{Q} \)), key (\( \mathbf{K} \)), and value (\( \mathbf{V} \)) representations, where query and key determine the relevance between tokens, and value encodes the feature information. 

The attention scores are computed using a scaled dot product:
\begin{equation}
    \text{Attention Scores} = \text{Softmax} \left( \frac{\mathbf{QK}^T}{\sqrt{d}} \right),
\end{equation}
where \( d \) is the key vector dimension. The weighted values are then processed to highlight significant frequency components while suppressing noise. The output is further refined through fully connected layers with ReLU activations to enhance feature extraction. For each \( \theta_p \), the output consists of two components, representing the real and imaginary parts of the signal. This output is then multiplied by the array manifold matrix to reconstruct the final signal.

This attention mechanism allows the network to adaptively focus on the most informative frequency components, improving robustness against sparsity-induced distortions and enhancing signal reconstruction.

\subsection{Dataset and Training Approach}\label{data}
We use a ULA with 20 elements and an inter-element spacing of half a wavelength to generate simulated single snapshot beam vectors for up to two targets. The radar field of view (FOV) is defined as $\boldsymbol{\phi}_{\rm FOV} = [-30^{\circ}, 30^{\circ}]$ and is uniformly discretized in the frequency domain into $64$ bins. The reflection coefficients \( s_k \) for each DOA source are randomly generated as complex numbers, where the amplitudes follow a uniform distribution in the range \([0.5, 1]\) and the phases are uniformly distributed within \([0, 2\pi]\). A total of $131,072$ signals are simulated for training.

The proposed FA-Net was trained for $500$ epochs with a batch size of $512$ using the Adam optimizer with a learning rate of $0.001$ and a mean squared error (MSE) loss function. The training setup included a maximum sparsity of $0.4$, with the SNR varying between $10$ dB and $30$ dB. The model was trained end-to-end to minimize the MSE loss and enhance signal reconstruction accuracy. All experiments were conducted using Python 3.8, PyTorch 1.10, and CUDA 11.1 on four Nvidia RTX A6000 GPUs.

%% file: section/4_performance.tex
\section{Performance Evaluation}\label{perf}
\begin{figure}[ht]
\centering
\includegraphics[width= 0.7\linewidth]{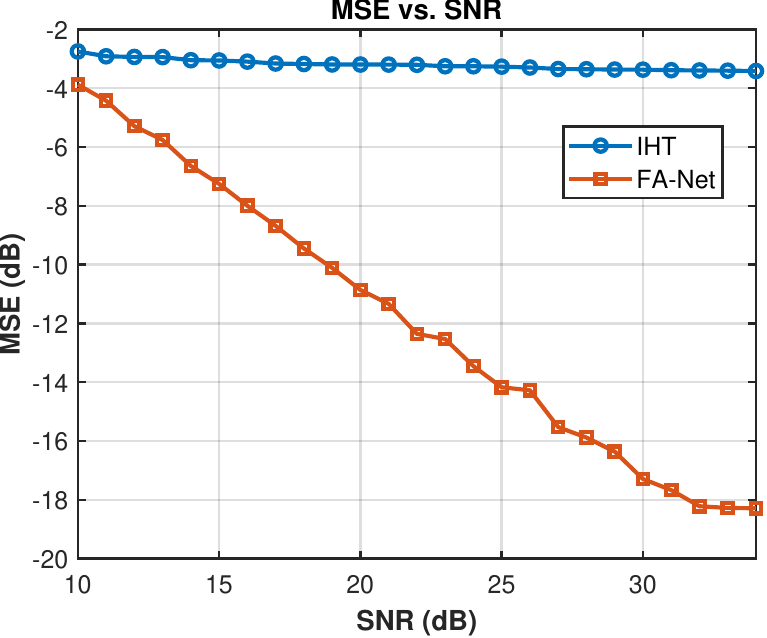}
\caption{Reconstruction error vs SNR.
}
\label{recon_err}
\vspace{-3mm}
\end{figure}

We evaluate FA-Net’s signal reconstruction using the MSE metric, comparing it to the iterative hard thresholding (IHT) algorithm \cite{cai2019fast}. Performance is assessed under $0.4$ sparsity, with fixed missing element counts but randomly varying positions across trials. A Monte Carlo approach with $5,000$ independent trials per SNR level ensures statistical reliability. Test signals contain two targets within the radar FOV. As shown in Fig. \ref{recon_err}, FA-Net outperforms IHT across all SNR levels. These results demonstrate FA-Net’s effectiveness in handling missing data and its superiority over conventional model-based approaches.

\begin{figure*}[ht]
    \centering
    \begin{subfigure}[b]{0.3\textwidth}
        \centering
        \includegraphics[width=\textwidth]{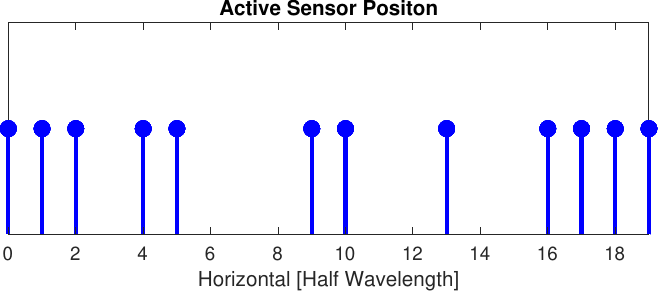}       
        \caption{}
        \vspace{1mm}
    \end{subfigure}
    \begin{subfigure}[b]{0.32\textwidth}
        \centering
        \includegraphics[width=\textwidth]{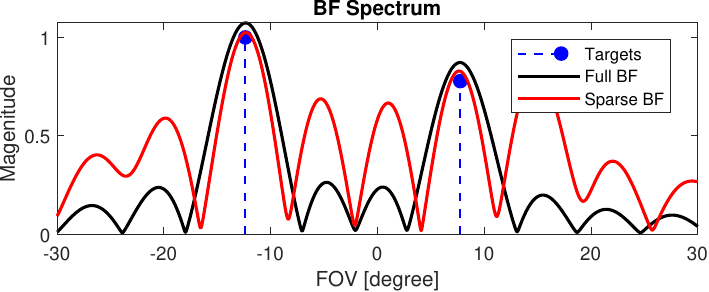}
        \caption{}
        \vspace{1mm}
    \end{subfigure}
    \begin{subfigure}[b]{0.32\textwidth}
        \centering
        \includegraphics[width=\textwidth]{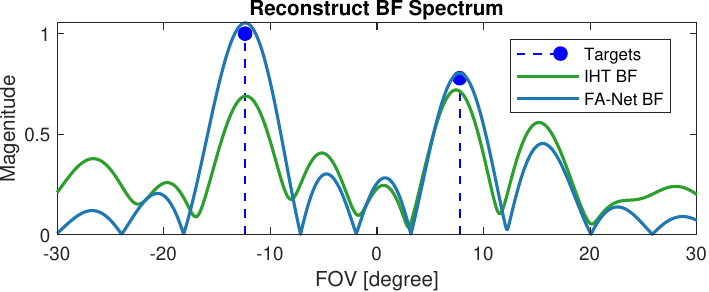}
        \caption{}
        \vspace{1mm}
    \end{subfigure}
    \begin{subfigure}[b]{0.3\textwidth}
        \centering
        \includegraphics[width=\textwidth]{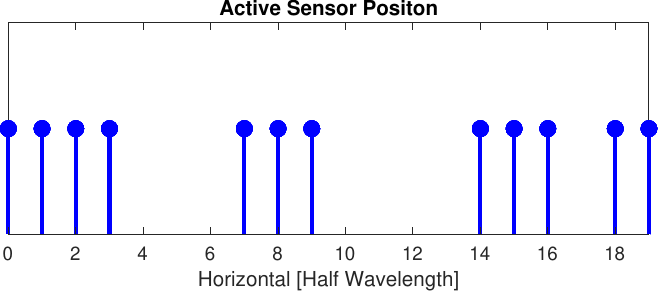}
        \caption{}
    \end{subfigure}
    \begin{subfigure}[b]{0.32\textwidth}
        \centering
        \includegraphics[width=\textwidth]{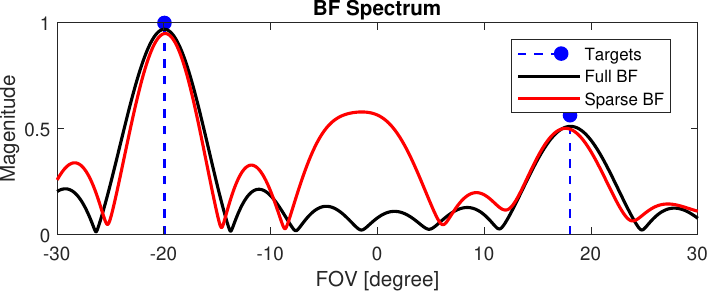}
        \caption{}
    \end{subfigure}
    \begin{subfigure}[b]{0.32\textwidth}
        \centering
        \includegraphics[width=\textwidth]{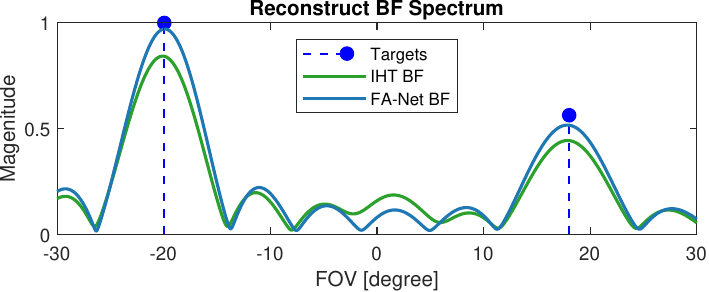}
        \caption{}
    \end{subfigure}
\vspace{-2mm}
    \caption{Signal reconstruction examples at 10 dB SNR (a–c) and 30 dB SNR (d–f).  
(a, d) Sparse array geometry. (b, e) Beamforming (BF) spectrum of the ground truth signal and sparse noise signal. (c, f) Reconstructed BF spectrum using the IHT algorithm and FA-Net.}
    \label{recon_exp}
    \vspace{-4mm}
\end{figure*}

To better illustrate the performance of FA-Net, we present reconstruction examples in Fig.~\ref{recon_exp}. Since the signals themselves are difficult to visualize directly, we use beamforming (BF) to obtain their spectral representations.

The first row of Fig.~\ref{recon_exp} (a-c) corresponds to a $10$ dB SNR scenario, while the second row (d-f) represents $30$ dB SNR, both with a sparsity level of $0.4$, where eight elements are missing. Subfigures (a) and (d) depict the sparse array geometry. Subfigures (b) and (e) compare the BF spectrum of the ground truth signal with that of the sparse, noise-affected signal, showing that sparsity and noise introduce higher sidelobes and spectral distortion. Subfigures (c) and (f) present the BF spectrum of the reconstructed signals using IHT and FA-Net. The results clearly demonstrate that FA-Net produces a more accurate BF spectrum, indicating superior reconstruction performance. These examples highlight FA-Net’s ability to adapt effectively to random sparse array geometries, outperforming traditional algorithms in signal reconstruction.

%% file: section/5_conclusion.tex
\section{Conclusions}
In summary, this paper presents a deep learning framework for sparse array signal reconstruction, enhancing the adaptability and robustness of automotive radar. By integrating sparse and noise signal augmentation, frequency-domain tokenization, and attention mechanisms, the deep learning model is generalized across diverse sparse geometries without retraining. Experimental results validated its effectiveness in suppressing sidelobe artifacts and improving signal fidelity.